\documentstyle[prb,multicol,aps,graphicx]{revtex}
\begin{document}
\draft 
\title{X-ray photoemission characterization of La$_{0.67}$(Ca$_{x}$Sr$_{1-x}$)$_{0.33}$MnO$_{3}$ 
films} 
\author{P. R. Broussard, V.M. Browning, and V. C. Cestone} 
\address{Naval Research Lab, 
Washington, DC 20375} 
\maketitle

\begin{abstract}
The Curie temperature and x-ray photoemission spectra of thin films of \nolinebreak
La$_{0.67}$(Ca$_{x}$Sr$_{1-x}$)$_{0.33}$MnO$_{3}$ (LCSMO) have been 
studied as a function of the Ca/Sr ratio.  The films were grown by 
off-axis cosputtering from individual targets of 
La$_{0.67}$Ca$_{0.33}$MnO$_{3}$ (LCMO) and 
La$_{0.67}$Sr$_{0.33}$MnO$_{3}$ (LSMO) onto (100) oriented NdGaO$_{3}$ 
substrates.  The films grow with a (100) orientation, with no other 
orientations observed by x-ray diffraction.  For the alloy mixtures, the Curie temperature, 
$T_{C}$, varies slowly as the Ca/Sr is decreased, remaining $\approx$ 
300 K, while for the LCMO and LSMO films $T_{C}$ is 260 and 330 K, 
respectively.  The Mn-O valence structure is composed of two dominant 
peaks, whose positions undergo a change as the Ca fraction is 
decreased.  The core lines behave as linear combinations of lines from 
pure LCMO and LSMO.  
\end{abstract}

\begin{multicols}{2}
\narrowtext	

\section{Introduction}
The effect of dopants on the photoemission from the lanthanum based 
manganese oxides has been an area of intense research activity, but 
usually the studies have focused on varying the Mn$^{3+}$/Mn$^{4+}$ 
ratio by changing the trivalent/divalent ion 
ratio.\cite{Saitoh1,Saitoh,Chainani,Taguchi}  While a great deal of 
work on studying magnetic, transport and structural properties of 
these compounds has looked at fixing the Mn$^{3+}$/Mn$^{4+}$ ratio and 
varying the size of the dopant atoms, little work has been done on 
photoemission studies.  In this paper we present an {\it in situ} 
x-ray photoemission spectroscopy (XPS) study of thin films of 
La$_{0.67}$(Ca$_{x}$Sr$_{1-x}$)$_{0.33}$MnO$_{3}$ (LCSMO) where the 
Mn$^{3+}$/Mn$^{4+}$ ratio is fixed while the tolerance factor, which 
is defined as $t=(d_{La-O})/\sqrt{2}(d_{Mn-O})$, is varied by the 
replacement of Ca with Sr.  A similar system was studied by Hwang et 
al.,\cite{Hwang} where the variation in the peak and Curie temperature 
for bulk samples of the CMR materials A$_{0.7}$A'$_{0.3}$MnO$_{3}$ 
(where A is a trivalent rare earth ion and A' is a divalent rare earth 
ion) were related to changes to the variation in the tolerance factor.  
In that work, the system 
La$_{0.7}$(Ca$_{x}$Sr$_{1-x}$)$_{0.3}$MnO$_{3}$ was studied, and 
exhibited a change in $t$ from $\approx$ 0.915 to 0.93 and Curie 
temperature from 250 to 365 K as x went from 1 to 0.

\section{Sample preparation and characterization}
The samples were grown by off-axis cosputtering onto (100) oriented neodymium 
gallate (NdGaO$_{3}$) substrates using composite targets 
of LCMO and LSMO material under similar 
conditions as our previous work.\cite{Broussard}   
The LCMO target was radio frequency (rf) sputtered and the LSMO target was 
direct current (dc) sputtered, giving deposition rates of $\approx$ 170-500 {\AA}/hr, with film thicknesses 
being typically 1000 {\AA}.  After deposition, the samples were cooled in 
100 Torr of oxygen, and when the samples had 
cooled to below 100 C, they were moved into a XPS analytical chamber. The
chamber pressure during XPS measurements was below 2 x 10$^{-9}$ Torr. 

The XPS spectra were taken at room temperature with a Vacuum 
Generators CLAM 100 analyzer using Mg K$\alpha$ radiation using the 
same conditions and analysis as in our earlier work.\cite{Broussard}  
Photoemission data for all the samples was collected around the Mn 2p 
doublet, the La 4d and 3d doublet, the O 1s, Sr 3d, Ca 2p, C 1s, Mn 3p 
lines, and the valence region.  In the following figures of XPS 
spectra, the core level spectra are shown along with their 
deconvolution into different Gaussian contributions.  

In addition to the XPS studies, the samples were characterized by 
standard $\theta-2\theta$ x-ray diffraction scans along the growth 
direction, atomic force microscopy, electrical resistivity 
measurements (using the van der Pauw method\cite{VdP}), and 
magnetization measurements at low fields using a Quantum Design SQUID 
Magnetometer.

\section{Results and Discussion}
\begin{figure}
\begin{center}
\includegraphics*[width=0.45\textwidth]{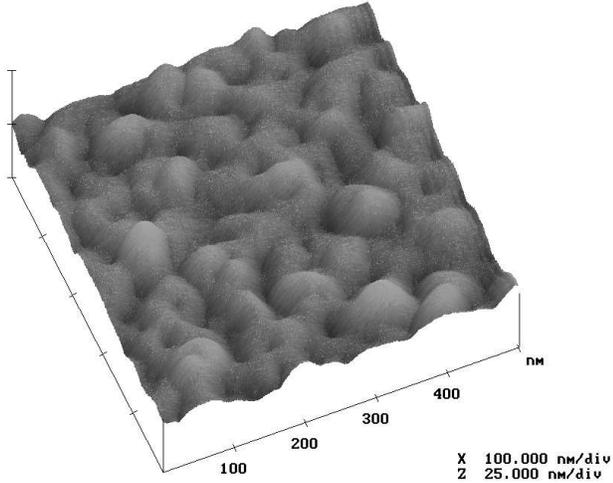}
\end{center}
\caption{Atomic force microscopy image for a LCSMO (61\% Ca) film grown on (100) 
NdGaO$_{3}$}
\label{AFM}
\end{figure}

\begin{figure}
\begin{center}
\includegraphics*[width=0.45\textwidth]{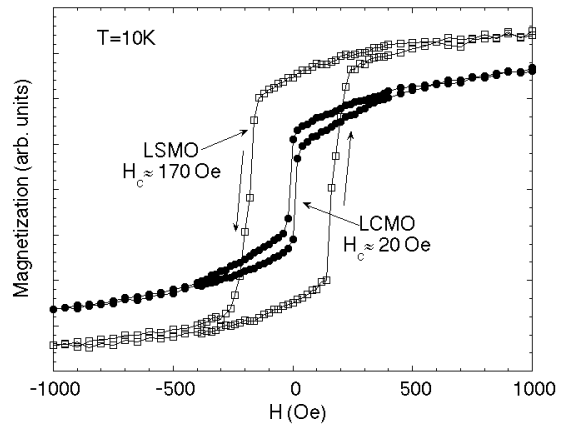}
\end{center}
\caption{Hysteresis loop at 10 K for pure LCMO and LSMO films}
\label{Hc}
\end{figure}

In Fig. \ref{AFM} we show an atomic force microscopy image for a 
LCSMO film with 61\% Ca fraction.  For pure LSMO and LCMO films we 
typically see surface roughness values of $\approx$ 16 {\AA}, while 
for the mixtures the value is typically 28 {\AA}.  Grain sizes are 
$\approx$ 500 {\AA} for the mixtures, while for pure LCMO and LSMO 
films they are closer to 1000 {\AA}.  Standard x-ray diffraction along 
the growth direction for LCMO films shows only the presence of peaks from 
NdGaO$_{3}$, indicating the excellent lattice match of the (100) 
oriented LCMO to (100) NdGaO$_{3}$.  This match is to be expected 
with the in-plane lattice constants of the pseudo-cubic unit cells of 
the two materials being 3.87 and 3.86 {\AA}, respectively.  For LSMO 
films, however, we find that the lattice match is not as good, with an 
out-of-plane lattice constant constant of 3.89 {\AA}.  This will give 
a strain in the films with low values of calcium doping.  This strain 
is seen in the increased coercive fields at 10 K for our LSMO films 
(170 Oe) compared to LCMO films (20 Oe), as seen in Fig. \ref{Hc}.

\begin{figure}
\begin{center}
\includegraphics*[width=0.45\textwidth]{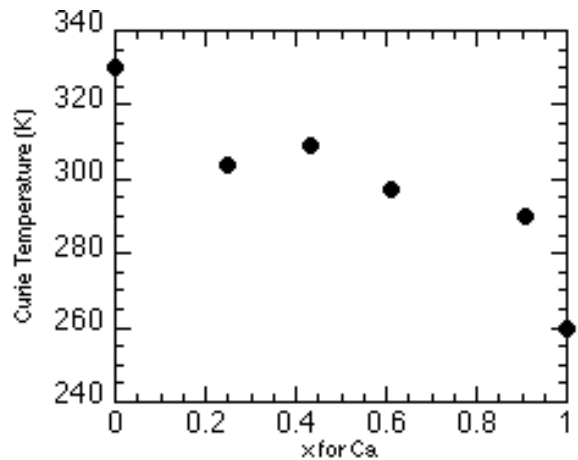}
\end{center}
\caption{Curie temperature vs. calcium fraction for LCSMO films 
grown on (100) NdGaO$_{3}$.}
\label{Curie}
\end{figure}

In Fig. \ref{Curie} we present the Curie 
temperatures ($T_{C}$) determined from magnetization data for the 
samples (taken at 400 Oe) as a function of the calcium fraction.  We 
do not see a smooth variation in the Curie temperature, as was seen 
in the bulk work.\cite{Hwang}  Instead, there seems to be a clustering 
of the values around 300 K, with a slight increase as the Sr content 
is increased, which would increase the tolerance factor.  We also observe a 
decrease in the Curie temperature for x=0.25 (where x is the calcium fraction in the 
sample) which is not expected.  We believe this drop is due to 
disorder in the samples, which is seen in the higher resistivity for 
this sample (not shown here) and would be consistent with 
increased strain in the films for lower values of x. 

\begin{figure}
\begin{center}
\includegraphics*[width=0.45\textwidth]{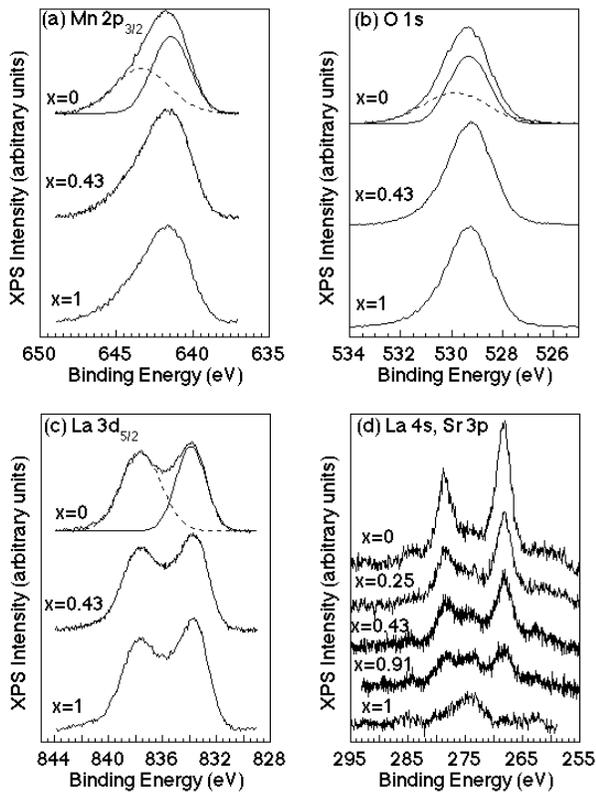}
\end{center}
\caption{XPS spectra and peak fit analysis of (a) Mn 2p$_{3/2}$, (b) 
O 1s, (c)  La 3d$_{5/2}$, and (d) La 4s, Sr 3p and C 1s lines.  Spectra are 
shown for various values of the calcium concentration, x.}
\label{XPS1}
\end{figure}

\begin{figure}
\begin{center}
\includegraphics*[width=0.45\textwidth]{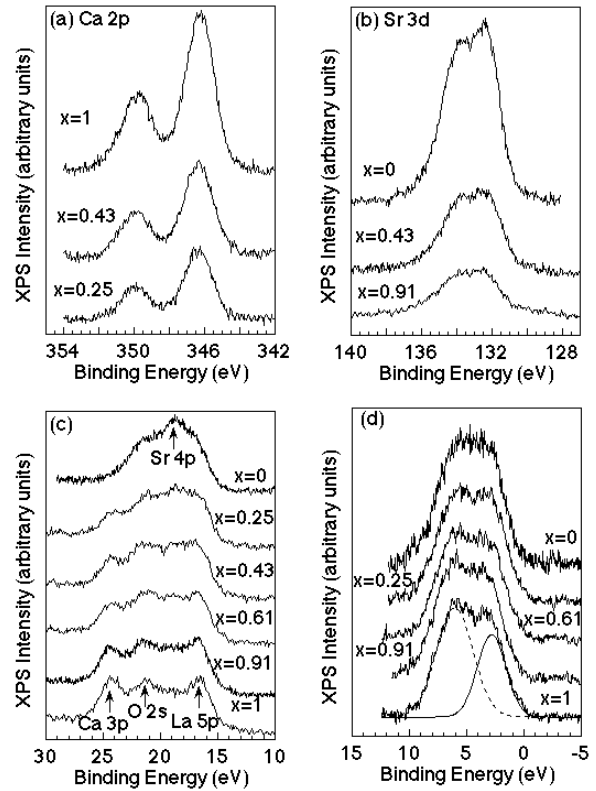}
\end{center}
\caption{Same as Figure \ref{XPS1} but for (a) Ca 2p, (b) Sr 3d, (c) the 
low energy core lines, and (d) the valence region.}
\label{XPS2}
\end{figure}

In Figs. \ref{XPS1} and \ref{XPS2}, we 
show scans around the location of the core lines and low energy 
region for the LCSMO films for different values of the calcium 
concentration, x.  The 
curves are offset for clarity.  We observe no indication for carbon 
for these {\it in-situ} samples, just as in our previous 
measurements.\cite{Broussard} For the core lines of Mn, O and La, we 
see little systematic variation as the Ca/Sr is varied (of the order 
of 0.1 eV), with the trend being that the peak positions for the core 
lines of LCMO having a slightly lower energy than LSMO.  Peak 
positions and widths are similar to that seen in earlier 
XPS studies of LCMO films.\cite{Broussard,Vasquez}  The lack of 
variation in peak position as the Ca/Sr ratio is changed is not surprising, in view of the 
XPS
work on La$_{1-x}$Sr$_{x}$MnO$_{3}$,\cite{Saitoh} which showed only slight 
variations ($\approx$ 0.3 eV in the metallic regime) in the core lines as the 
Sr fraction changed.  Scans of 
the Ca and Sr core lines show the expected variation in intensity as 
the Ca/Sr ratio is varied, but again, there is no systematic 
variation in the peak positions, which for the Sr 3d$_{5/2}$ is 132.0 
eV.  The peak ratios are similar to that seen in our earlier study on 
LCMO,\cite{Broussard} implying that for these systems the terminating 
layer is MnO$_{2}$

In Fig. \ref{XPS2}d we show the valence structure for the samples as a 
function of the calcium fraction.  As the calcium fraction decreases, 
and the samples go from insulating to metallic at room temperature, we 
observe no change near the Fermi edge, which is similar to that seen 
in
studies on La$_{1-x}$Sr$_{x}$MnO$_{3}$\cite{Saitoh,Chainani}.  In 
previous studies of the valence band of LSMO using 
XPS,\cite{Chainani}, a peak at 5.8 eV was observed, which did not 
change position as the Sr fraction was increased.  Instead, a 
reduction in intensity near 3-4 eV was observed, which was interpreted 
as being due to changes in doping.  For our system, we are not 
changing the filling of the $e_{g\uparrow}$ band, since we are keeping 
the doping fixed.  However we are changing the overlap integral as the 
value of t is varied.  We also find we can fit the valence structure 
to a doublet structure, as shown in Fig.  \ref{XPS2}.  As the samples 
go from pure Sr to pure Ca doping, we find the low and high energy 
peaks change from 2.7 to 2.9 eV, and 5.6 to 6.0 eV, respectively.  
This trend in peak positions is opposite that seen for the core lines, 
which tend to lower their energy in going from LSMO to LCMO. We do not 
observe any systematic changes in spectral weight in the valence 
region as the samples go from pure LSMO to LCMO.

\section{Conclusions}
Thin film alloy mixtures of (100) oriented 
La$_{0.67}$(Ca$_{x}$Sr$_{1-x}$)$_{0.33}$MnO$_{3}$ have been grown and 
have somewhat rougher surfaces and smaller grain sizes 
than seen for pure LSMO or LCMO. The Curie temperature for the films 
does not follow the expected smooth variation as the calcium fraction 
is changed, which we interpret as being due to disorder.  We believe 
this disorder is caused by strain in the films which is observed in 
both X-ray diffraction and coercive field measurements.  The core 
lines in XPS measurements behave as linear combinations of individual 
LCMO and LSMO films, with no significant change in position as the 
Ca/Sr ratio is varied.  The intensity ratios of the peaks is similar 
to previous work indicating that the terminating layer for the films 
is MnO$_{2}$.  The low energy valence structure exhibits a 
doublet character, whose peak positions decrease as the Ca fraction is 
reduced.  We interpret this change as being due to variations in the 
overlap integral between the Mn 3d and O 2p orbitals.

\section{Acknowledgments}
We would like to gratefully acknowledge the assistance of Michael 
Miller for the AFM measurements and Andrew Patton in the 
production of the films.

\end{multicols}

\begin{references}

\bibitem{Saitoh1}
T. Saitoh, A. Sekiyama, K. Kobayashi, T. Mizokawa, A. Fujimori, D.D. 
Sarma, Y. 
Takeda, and M. Takano, Phys.  Rev. B {\bf 56}, 8836 (1997).

\bibitem{Saitoh}
T. Saitoh, A. E. Bocquet, T. Mizokawa, H. Namateme, M. Abbate, Y. 
Takeda, and M. Takano, Phys.  Rev. B {\bf 51}, 13942 (1995).

\bibitem{Chainani}
A. Chainani, M. Matthew and D.D. Sarma, Phys.  Rev.  {\bf B47}, 15397 
(1993).

\bibitem{Taguchi}
H. Taguchi and M. Shimada, J. Sol.  St.  Chem.  {\bf 67}, 37 (1987).

\bibitem{Hwang}
H.Y. Hwang, S-W. Cheong, P.G. Radaelli, M. Marezio, and B. Batlogg, 
Phys. Rev. Lett. {\bf 75}, 914 (1995).

\bibitem{Broussard}
P.R. Broussard, S.Q. Qadri, V.M. Browning and V.C. Cestone, Appl.  
Surf.  Sci {\bf 115}, 80 (1997).

\bibitem{Shirley}
 D. A. Shirley, Phys.  Rev.  {\bf B5}, 4709 (1972).

\bibitem{VdP}
L. J. van der Pauw, Phillips Res.  Rep.  {\bf 13}, 1 (1958).

\bibitem{Vasquez}
R.P. Vasquez, Phys. Rev. B {\bf 54}, 14938 (1996).

\end{references}
\end{document}